# Ranking of nodes of networks taking into account the power function of its weight of connections

Soboliev A.M. [1], Lande D.V. [2]

[1] Post-graduate student of the Institute for Special Communications and Information Protection of NTUU «KPI named. Igor Sikorsky»

[2] Institute for Special Communications and Information Protection of NTUU «KPI named. Igor Sikorsky»

artem1sobolev@gmail.com, dwlande@gmail.com

**Abstract.** To rank nodes in quasi-hierarchical networks of social nature, it is necessary to carry out a detailed analysis of the network and evaluate the results obtained according to all the given criteria and identify the most influential nodes. Existing ranking algorithms in the overwhelming majority estimate such networks in general, which does not allow to clearly determine the influence of nodes among themselves. In the course of the study, an analysis of the results of known algorithms for ranking the nodes of HITS, PageRank and compares the obtained data with the expert evaluation of the network. For the effective analysis of quasi-hierarchical networks, the basic algorithm of HITS is modified, which allows to evaluate and rank nodes according to the given criteria (the number of input and output links among themselves), which corresponds to the results of expert evaluation. It is shown that the received method in some cases provides results that correspond to the real social relation, and the indexes of the authorship of the nodes - pre-assigned social roles.

## 1. Introduction

At present, when social networks are a component of society and are one of the mechanisms of influence on it, it is necessary to clearly understand all the possible risks that arise from it and be prepared for serious consequences in case of careless attitude to this situation. The expression "social networks" not only relates to social networks from the Internet, which are used as a means of communication, but also all communications that occur between people (social entities) for the transmission of information.

Also, it should be noted that the social network should be understood as a network, the nodes of which are social subjects (social subject), and contacts - contacts that are accompanied by the exchange of information between them.

By definition, a quasi-hierarchical network should be considered a network that is close to the hierarchical network, but its hierarchy is violated by a small number of additional links.

## 2. The purpose of the work

The purpose of this work is to describe and evaluate the effectiveness of the proposed algorithm for ranking nodes of quasi-hierarchical networks of social nature, based on the modification of the known HITS algorithm in comparison with other algorithms for ranking nodes in the networks.

When ranking a social network, one must understand the process of organizing nodes (social entities) on a specific basis, which allows one to determine the influence of the given nodes among themselves.

For visual representation, this network is represented in the form of a directed graph, in which nodes are markers of subjects (social network profiles, telephone numbers, electronic mailboxes, etc.), and the links are directed to the links that occur between them.

For a mathematical representation, such a graph is depicted in the form of a matrix of adjacent nodes and the value in the cell of the matrix reflects the number of directed bonds.

To determine the impact of the nodes and to demonstrate the use of the Hyperlink-Induced Topic Search (HITS), consider a conventional hierarchical network of 6 nodes (Figure 1), to which each node will calculate the coefficients of auth and hub.

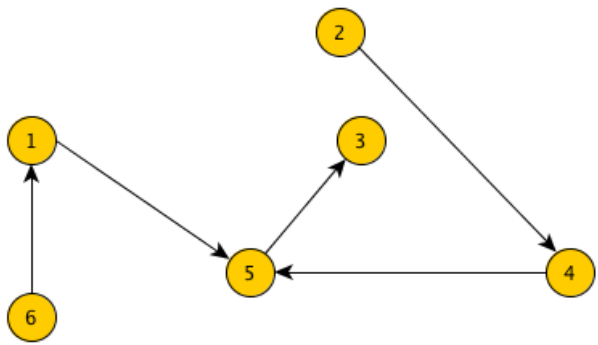

**Fig. 1.** Hierarchical network

Let us show the connections between nodes in the form of an adjacency matrix:

$$L = \begin{bmatrix} 0 & 0 & 0 & 0 & 1 & 0 \\ 0 & 0 & 0 & 1 & 0 & 0 \\ 0 & 0 & 0 & 0 & 0 & 0 \\ 0 & 0 & 0 & 0 & 1 & 0 \\ 0 & 0 & 1 & 0 & 0 & 0 \\ 1 & 0 & 0 & 0 & 0 & 0 \end{bmatrix},$$

After the transposition we obtain the following matrix:

$$L^T = \begin{bmatrix} 0 & 0 & 0 & 0 & 0 & 1 \\ 0 & 0 & 0 & 0 & 0 & 0 \\ 0 & 0 & 0 & 0 & 1 & 0 \\ 0 & 1 & 0 & 0 & 0 & 0 \\ 1 & 0 & 0 & 1 & 0 & 0 \\ 0 & 0 & 0 & 0 & 0 & 0 \end{bmatrix},$$

To calculate auth and hub of the basic HITS algorithm for data presented in the form of matrices, we use the following iterative formulas [Amy N. Langville, Carl D. Meyer, 2006]:

$$auth^{(k)} = L^T \cdot hub^{(k-1)},$$

$$hub^{(k)} = L \cdot auth^{(k)},$$

$$k = k + 1,$$

Where $auth^{(k)}, hub^{(k)}$ - vectors containing coefficients of auth and hub for each iteration k, over a certain number of repetitions.
We modify the above-mentioned formulas to ensure the ability to count auth and hub indices independently of each other [Amy N. Langville, Carl D. Meyer, 2006]:

$$auth^{(k)} = L^T L auth^{(k-1)},$$

$$hub^{(k)} = L L^T hub^{(k-1)},$$

$$k = k + 1,$$

For initialization of the formulas under the initial matrices $auth^{(0)}$ and $hub^{(0)}$ we use single vectors:

$$L^T L = \begin{bmatrix} 0 & 0 & 0 & 0 & 0 & 1 \\ 0 & 0 & 0 & 0 & 0 & 0 \\ 0 & 0 & 0 & 0 & 1 & 0 \\ 0 & 1 & 0 & 0 & 0 & 0 \\ 1 & 0 & 0 & 1 & 0 & 0 \\ 0 & 0 & 0 & 0 & 0 & 0 \end{bmatrix} \cdot \begin{bmatrix} 0 & 0 & 0 & 0 & 1 & 0 \\ 0 & 0 & 0 & 1 & 0 & 0 \\ 0 & 0 & 0 & 0 & 0 & 0 \\ 0 & 0 & 0 & 0 & 1 & 0 \\ 0 & 0 & 1 & 0 & 0 & 0 \\ 1 & 0 & 0 & 0 & 0 & 0 \end{bmatrix} = \begin{bmatrix} 1 & 0 & 0 & 0 & 0 & 0 \\ 0 & 0 & 0 & 0 & 0 & 0 \\ 0 & 0 & 1 & 0 & 0 & 0 \\ 0 & 0 & 0 & 1 & 0 & 0 \\ 0 & 0 & 0 & 0 & 2 & 0 \\ 0 & 0 & 0 & 0 & 0 & 0 \end{bmatrix},$$

$$L^T L = \begin{bmatrix} 0 & 0 & 0 & 0 & 0 & 1 \\ 0 & 0 & 0 & 0 & 0 & 0 \\ 0 & 0 & 0 & 0 & 1 & 0 \\ 0 & 1 & 0 & 0 & 0 & 0 \\ 1 & 0 & 0 & 1 & 0 & 0 \\ 0 & 0 & 0 & 0 & 0 & 0 \end{bmatrix} \cdot \begin{bmatrix} 0 & 0 & 0 & 0 & 1 & 0 \\ 0 & 0 & 0 & 1 & 0 & 0 \\ 0 & 0 & 0 & 0 & 0 & 0 \\ 0 & 0 & 0 & 0 & 1 & 0 \\ 0 & 0 & 1 & 0 & 0 & 0 \\ 1 & 0 & 0 & 0 & 0 & 0 \end{bmatrix} = \begin{bmatrix} 1 & 0 & 0 & 0 & 0 & 0 \\ 0 & 0 & 0 & 0 & 0 & 0 \\ 0 & 0 & 1 & 0 & 0 & 0 \\ 0 & 0 & 0 & 1 & 0 & 0 \\ 0 & 0 & 0 & 0 & 2 & 0 \\ 0 & 0 & 0 & 0 & 0 & 0 \end{bmatrix},$$

$$k = 1,$$

$$auth(k) = \begin{bmatrix} 1 & 0 & 0 & 0 & 0 & 0 \\ 0 & 0 & 0 & 0 & 0 & 0 \\ 0 & 0 & 1 & 0 & 0 & 0 \\ 0 & 0 & 0 & 1 & 0 & 0 \\ 0 & 0 & 0 & 0 & 2 & 0 \\ 0 & 0 & 0 & 0 & 0 & 0 \end{bmatrix} \cdot \begin{bmatrix} 1 \\ 1 \\ 1 \\ 1 \\ 1 \\ 1 \end{bmatrix} = \begin{bmatrix} 1 \\ 0 \\ 1 \\ 1 \\ 2 \\ 0 \end{bmatrix},$$

$$hub(k) = \begin{bmatrix} 1 & 0 & 0 & 1 & 0 & 0 \\ 0 & 1 & 0 & 0 & 0 & 0 \\ 0 & 0 & 0 & 0 & 0 & 0 \\ 1 & 0 & 0 & 1 & 0 & 0 \\ 0 & 0 & 0 & 0 & 1 & 0 \\ 0 & 0 & 0 & 0 & 0 & 1 \end{bmatrix} \cdot \begin{bmatrix} 1 \\ 1 \\ 1 \\ 1 \\ 1 \\ 1 \end{bmatrix} = \begin{bmatrix} 2 \\ 1 \\ 0 \\ 2 \\ 1 \\ 1 \end{bmatrix},$$

From the results of nodes ranking by the basic HITS algorithm we will conclude that the fifth node contains the most relevant information, and nodes numbered 2 and 4 are the most portable, that is, they have a link to the node that contains the most useful information for a given search.

The considered graph consists of a small number of links, and in this case, it is possible to visualize it and make the assumption that the most influential in this network is node number 5, since it has the largest number of input and output links.
According to the HITS algorithm, two nodes are calculated for each network node:

- mediation indicator (hub);
- index of authorship (auth).

The mediation rate of the *hub* $A_i$ of the node $A_j$ of the network, consisting of n nodes, is equal to the sum of the values of the authorship of the nodes to which it refers:

$$auth\ A_i = \sum_{j \to i} hub\ A_j,$$

The given formulas give a fast convergent iterative algorithm for calculating hub and auth for all nodes in the network.

At a time when it comes to more complex networks (large-scale networks), where the number of nodes can be more than 1000 (large organizations, etc.), it is almost impossible without the special algorithms of the assumptions of the social actors that act as leaders of different rank [Lande D.V., Nechaev A.O., 2015].

Scale-free networks have the characteristic such as the distribution of nodes, which is defined as the probability that a node has a degree (the node's degree is the number of edges associated with that node). It is the networks with the indicative distribution of node degrees called scale-free, which are most often observed in real-life large networks.

It should be noted that the number of links in the quasi-hierarchical networks of social nature is equal to the number of nodes and this statement can be represented in the following mathematical form:

$$N_i = O(N_j),$$

Where $N_i$- the number of connections in the network, $N_j$- the number of nodes in the network.

If, when considering the quasi-hierarchical networks of social nature, there is a clear (and also significant) one-way communication, then even there is no question as to which of the pairs of nodes has a higher rank.

In these networks, there are three general variants of relations for a pair of nodes:

- The subordinate leader;
- Knots of the same rank;
- Nodes of the lower rank have a direct connection with nodes of a higher rank.

In the case where subject *A* is the head of subject *B* in the social network it is proposed to consider as the aggregate weight of factors, one of the criteria of which is the time indicator.

It should be noted that the head is chaotic with the subordinates, if necessary, the task, and the subordinate with a certain regular frequency reports on the implementation of the completed assignment to the head.

In assessing the characteristics of the number of links, there is the fact that the head is more often associated with subordinates, that is, the weight of the outgoing communication from the head must exceed the weight of the incoming connection [Lande D.V., Nechaev A.O., 2015].

It should be noted that only the generalized and most widespread template for subordinate and manager contacts is considered. In real work, this template can be supplemented in some cases by other specific characteristics, which will increase the effectiveness of the weight estimation of pairs of nodes contacts [Liu Y.Y., Jean-Jacques Slotine J.J., Barabasi A.L., 2012].

Occasionally, when a lower-level employee assists a higher-ranking employee in work processes, or a higher-ranking manager gives the employee a much lower level of hierarchy, this leads to the emergence of non-hierarchical links in such a network. Such cases, as a rule, are not much to consider the network as quasi-hierarchical. Meanwhile, the use of the basic HITS algorithm for ranking nodes in the above mentioned network may not provide the desired result, because even a weak and insignificant connection (if not taking into account its weight) can greatly affect the portability and authorship of nodes [Bargh JA, Chen M., 1996].

You can use PageRank algorithms to rank nodes in social networks:

$$PR(A) = (1 + d) + d \sum_{i=1}^{n} \frac{PR(T_i)}{C(T_i)},$$

Where *PR(A)* - the PageRank weight for page *A*, *d* - the attenuation factor *PR(T$_i$)*- the PageRank weight of the page that points to page *A*, *C(T$_i$)*- the number of links from this page, *n* - the pages linking to page *A*.

## 3.　　Presentation of the main results

For the mathematical realization of the iterative principle of ranking nodes in large quasi-hierarchical networks of social nature should be noted that the most influential on the network entity is the one who has the largest number of outbound connections with other network entities.

To demonstrate and determine the effectiveness of nodes ranking, consider the quasi-hierarchical networks of social nature, which consists of 14 subjects (Figure 2), between which certain links are known to be distributed, and we use the above-mentioned algorithms and compare the results obtained with each other.

When ranking the network algorithm HITS in the final result we get two parameters: auth and hub, which does not allow to reject the network at the same time according to two criteria. To solve this question, you must use a metric that combines information about auth and hub of our algorithm. It is by such a metric that we use the F-measure:

$$F(A_j) = \frac{2}{\frac{1}{auth\ A_j} + \frac{1}{hubA_j}},$$

Where $F(A_j)$ - F-measure of node $A_j$ network, *hub A$_i$* - mediation indicator *auth A$_i$* - index of authorship.

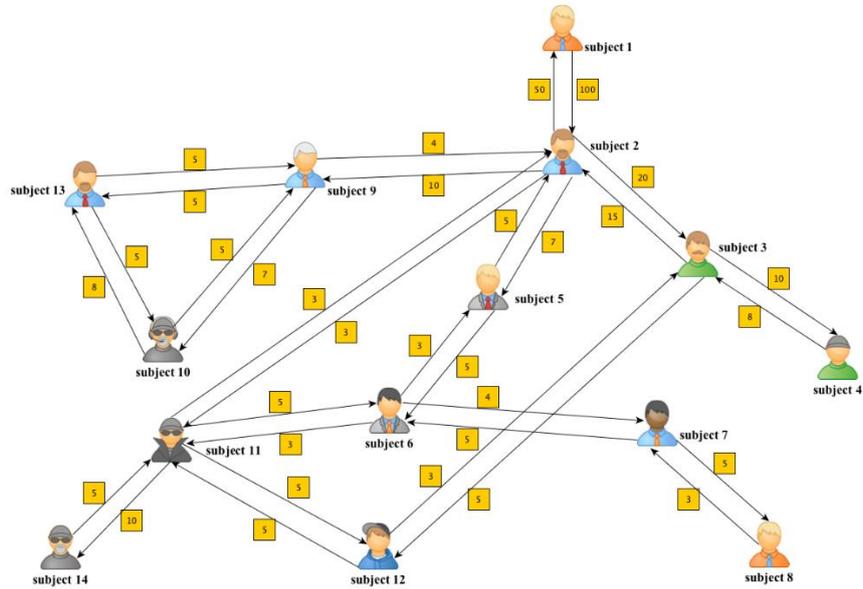

**Fig. 2.** Quasi-hierarchical network of social character

**Table 1.** Results of ranking of network nodes using PageRank algorithms and the basic HITS algorithm and expert assessment

| Expert evaluation (No. node) | No. node | PageRank | No. node | HITS auth hub F |
|---|---|---|---|---|
| 1 | 2 | 0.167483 | 2 | 0.153005 |
| 2 | 11 | 0.106627 | 11 | 0.128415 |
| 3 | 1 | 0.089799 | 9 | 0.056011 |
| 9 | 3 | 0.085644 | 3 | 0.092896 |
| 5 | 9 | 0.076856 | 5 | 0.094262 |
| 11 | 6 | 0.073306 | 6 | 0.086066 |
| 4 | 10 | 0.067540 | 12 | 0.050546 |
| 12 | 13 | 0.066458 | 1 | 0.050546 |
| 10 | 7 | 0.063547 | 10 | 0.036885 |
| 13 | 14 | 0.050122 | 13 | 0.043716 |
| 6 | 12 | 0.042552 | 14 | 0.075136 |
| 14 | 8 | 0.037723 | 7 | 0.013661 |
| 7 | 5 | 0.037363 | 4 | 0.087431 |
| 8 | 4 | 0.034981 | 8 | 0.031421 |

Having analyzed the obtained results of network ranking (Table 1), it is evident that the parameters of auth and hub are equal, since in the quasi-hierarchical networks of social nature basically there are feedback links between nodes and the basic algorithm of HITS does not provide the necessary result of ranking such networks. Also, comparing the results of the algorithms of HITS and PageRank with expert assessment data, it can be argued that in the basic form, the above-mentioned algorithms are ineffective for ranking nodes in the quasi-hierarchical networks of social nature.

PageRank takes into account only inbound links and does not distinguish between such an entity as an intermediary, which has a significant impact on the overall result, since minor links in the network can radically change it.

To adapt the method, it is proposed to use the modified HITS algorithm and, in calculating each indicator of the network nodes, take into account the weight of the ribs. At the same time, the uneven distribution of weight values may affect the reliability of the result, therefore their value must be reduced by multiplying by some monotonically increasing function, less steep than linear. Within the framework of the proposed algorithm, as a function, we use the modified HITS algorithm and modify the formulas by summing up the weight of the links between the nodes and from this follows:

$$hub\ A_i = \sum_{j \leftarrow i} auth\ A_j E_{ij}^a,$$

$$auth\ A_i = \sum_{j=1}^{n} hub\ A_j \cdot E_{ij}^a,$$

Where hub $A_j$ is the mediation rate of the node $auth\ A_i$ of the network, auth $A_j$ is the index of authorship of the node $A_j$, $E_{ij}$ is the weight of the links between the nodes $A_i$ and $A_j$ (respectively, between the nodes $A_i$ and $A_j$)), $a$ is the degree.

Table 2. Results of ranking of network nodes with modified PageRank algorithm and expert assessment

| Expert evaluation (No. node) | No. node | HITS $a = 1$ | | |
|---|---|---|---|---|
| | | auth | hub | F |
| 1 | 1 | 0.100370 | 0.731910 | 0.176532 |
| 2 | 2 | 0.787807 | 0.059213 | 0.110147 |
| 3 | 3 | 0.041181 | 0.111308 | 0.060119 |
| 9 | 9 | 0.020522 | 0.029625 | 0.024247 |
| 5 | 5 | 0.014123 | 0.036793 | 0.020411 |
| 11 | 11 | 0.006428 | 0.022866 | 0.010036 |
| 4 | 4 | 0.011803 | 0.003188 | 0.005020 |
| 12 | 12 | 0.007220 | 0.001528 | 0.002522 |
| 10 | 10 | 0.002461 | 0.001232 | 0.001642 |
| 13 | 13 | 0.001983 | 0.001163 | 0.001466 |
| 6 | 6 | 0.003320 | 0.000620 | 0.001044 |
| 14 | 14 | 0.002637 | 0.000333 | 0.000591 |
| 7 | 7 | 0.000097 | 0.000212 | 0.000134 |
| 8 | 8 | 0.000048 | 0.000008 | 0.000014 |

Having evaluated the results of the modified HITS algorithm at ($a = 1$) (Table 2) with expert assessment data, it can be argued that they coincide with each other and the modified HITS method is effective in the overall assessment in the quasi-hierarchical networks of social nature. For a detailed assessment of the network, it is necessary to analyze the obtained results of auth and hub with experts data, since even minor relationships can significantly affect the ranking of results and the identification of the most significant nodes in the network.

**Table 3.** Results of ranking of network nodes with modified PageRank algorithm and expert assessment.

| Expert evaluation (No. node) | No. node (auth) | No. node (hub) |
|---|---|---|
| 1 | 2 | 1 |
| 2 | 1 | 3 |
| 3 | 3 | 2 |
| 9 | 9 | 5 |
| 5 | 5 | 9 |
| 11 | 4 | 11 |
| 4 | 12 | 4 |
| 12 | 11 | 12 |
| 10 | 6 | 10 |
| 13 | 14 | 13 |
| 6 | 10 | 6 |
| 14 | 13 | 14 |
| 7 | 7 | 7 |
| 8 | 8 | 8 |

After studying the results (Table 3), it can be argued that the numbers of the most significant redirected nodes according to auth and hub are different from the results of the expert assessment and require the correction of degree a in the above formula. To evaluate the network, use the parameter $a = \frac{2}{3}, \frac{2}{5}$ and carry out a repeated study.

**Table 4.** Results of ranking of network nodes with modified PageRank algorithms and expert assessment.

| Expert evaluation (No. node) | HITS, $a = \frac{2}{3}$ | | | HITS, $a = \frac{2}{5}$ | | |
|---|---|---|---|---|---|---|
| | No. node (auth) | No. node (hub) | F | No. node (auth) | No. node (hub) | F |
| 1  | 2  | 1  | 1  | 1  | 1  | 1  |
| 2  | 1  | 3  | 2  | 2  | 3  | 3  |
| 3  | 3  | 2  | 3  | 3  | 2  | 2  |
| 9  | 9  | 5  | 9  | 9  | 5  | 9  |
| 5  | 5  | 9  | 5  | 5  | 9  | 5  |
| 11 | 11 | 11 | 11 | 11 | 11 | 11 |
| 4  | 4  | 4  | 4  | 12 | 10 | 4  |
| 12 | 12 | 10 | 12 | 4  | 4  | 12 |
| 10 | 6  | 12 | 10 | 6  | 13 | 10 |
| 13 | 14 | 13 | 13 | 14 | 12 | 13 |
| 6  | 10 | 6  | 6  | 13 | 6  | 6  |
| 14 | 13 | 14 | 14 | 10 | 14 | 14 |
| 7  | 7  | 7  | 7  | 7  | 7  | 7  |
| 8  | 8  | 8  | 8  | 8  | 8  | 8  |

In the process of studying the results obtained in table 4 modified HITS algorithm, there is a significant difference in the calculated parameters and it is determined that when using the parameter $a = \frac{2}{3}$, the modified method in a number of cases provides results that correspond to the real social relation and the auth is sufficiently accurate to show the hierarchical dependence of the subjects among themselves on the network.

It should be noted that this modified algorithm with the value of the parameter $a = \frac{2}{3}$ was used to study twenty different quasi-hierarchical networks, and the obtained results of the indices corresponded to the given reality.

## Conclusions

The concept of quasi-hierarchical networks of social character is introduced in the work and the nature of connections in such networks is investigated.

The research of the efficiency of the basic and modified HITS algorithms for ranking of nodes of quasi-hierarchical networks of social character in comparison with the PageRank algorithm on the given network was conducted.

In practical application it is shown that the modified algorithm provides in some cases the results that correspond to the real social relation, and the auth of the nodes - their hierarchical importance among themselves.